\documentclass[final,3p,times,twocolumn]{elsarticle}
\usepackage{epsf,epsfig,graphicx}

\journal{Astroparticle Physics}

\begin{document}

\begin{frontmatter}

\title{Atmospheric ionization and cosmic rays:\\studies and measurements before 1912}

\author[Udine,Lisboa]{Alessandro De Angelis}

\address[Udine]{INFN and Universit\`a di Udine, Via delle Scienze 206, I-33100 Udine, Italy}
\address[Lisboa]{LIP/IST Lisboa, Portugal}

\begin{abstract}
The discovery of cosmic rays, a milestone in science, was based on the work by scientists in Europe and the New World and took place during a period characterised by nationalism and lack of communication. Many scientists that took part in this research a century ago were intrigued by the penetrating radiation and tried to understand the origin of it. Several important contributions to the discovery of the origin of cosmic rays have been forgotten; historical, political and personal facts might have contributed to their substantial disappearance from the history of science.\end{abstract}


\end{frontmatter}

\section{The spontaneous discharge of electroscopes}

\begin{figure}
\begin{center}
\resizebox{0.82\columnwidth}{!}{\includegraphics{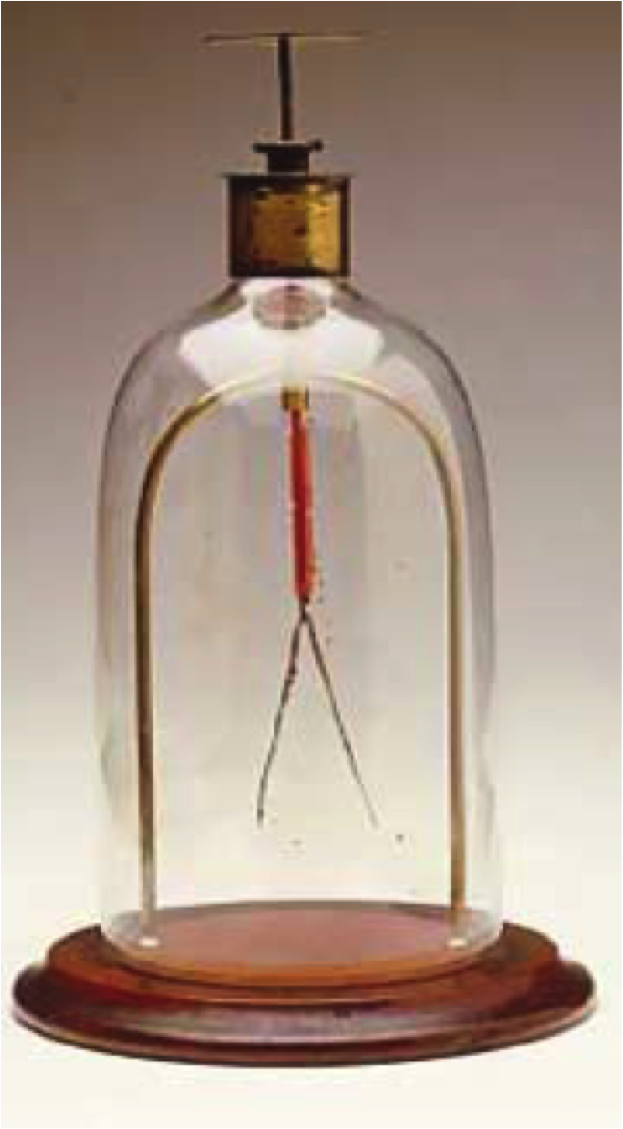} }
\end{center}
\caption{An electroscope of the end of XVIII century.}
\label{fig:electroscope}       
\end{figure}

A typical electroscope, in the configuration which was invented at the end of the XVIII century (Figure \ref{fig:electroscope}), consists of a vertical metal rod from the end of which hang two gold leaves. A disk or ball terminal is attached to the top of the rod, where the charge to be tested is applied. To protect the gold leaves from drafts of air they are enclosed in a glass bottle. The gold leaves repel, and thus diverge, when the rod is charged.

 One could think at first glance that, if  isolation were perfect, an electroscope should always maintain its charge.  An unexpected result came from the first experiments related to electricity by Charles-Augustin de Coulomb, officer of the French army and member of the Acad\'emie des sciences.  Coulomb was surprised in finding, around 1785  \cite{Cou1785},  that electroscopes can spontaneously discharge
by the action of the air and not by defective insulation. He published this result in his famous ``M\'emoires sur l'\'electricit\'e et le magn\'etisme''.

\begin{figure}
\begin{center}
\resizebox{0.82\columnwidth}{!}{\includegraphics{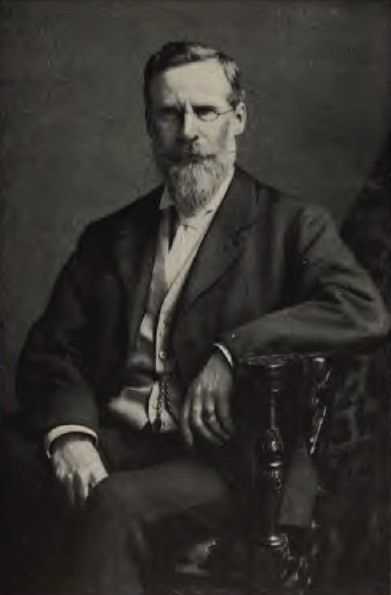} }
\end{center}
\caption{Sir William Crookes (1832-1919) -- source: wikimedia commons.}
\label{fig:crookes}       
\end{figure}

 Also Michael Faraday  addressed the  problem around 1835 \cite{Far1835}, confirming with greater accuracy the results by Coulomb. In the meantime the electroscope was improved by William Thomson, then Lord Kelvin; Crookes \cite{Cro1879} (Figure \ref{fig:crookes}) could  measure in 1879  that the speed of discharge of an
electroscope decreased when the air pressure was reduced. It became therefore clear that the direct cause of the discharge of the electroscope should be the ionization of the air contained in the instrument itself. But what was the cause of this ionization?
 
The explanation of the phenomenon of spontaneous discharge
came in the beginning of the 20th century and paved the way to one of mankind's
revolutionary scientific discoveries \cite{dea}: cosmic rays.

\section{The puzzle of atmospheric ionization}

The study of the rate of discharge of an electroscope required a rather sophisticated experimental technology; fortunately this type of measurement was very popular since the
late eighteenth century, as related to issues concerning atmospheric electricity, and 
ultimately meteorology. The technique was also developed  in the 
United States, Canada, Italy, Germany, and particularly in Austria. In most
cases these studies were financed  thanks to the possible interest for agriculture and military science, two areas
which would have greatly benefited from the possibility that humans were able to influence
the weather thanks to electricity. 


Franz Exner, whose school in Vienna was
rewarded by several Nobel prizes \cite{craw}, not only further perfected the electroscope
improving the tools of Lord Kelvin, but also managed to attract many good students -- for example the future Nobel prize-winner
Schr\"odinger, who became interested  to physics thanks to the study of ionization of the atmosphere.

\begin{figure}
\begin{center}
\resizebox{\columnwidth}{!}{ \includegraphics{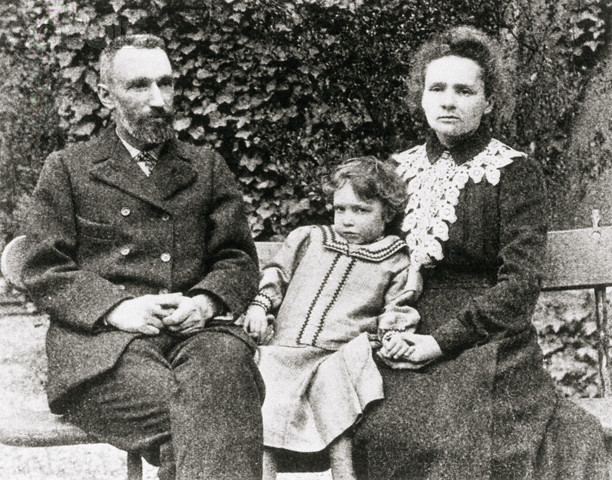} }
\end{center}
\caption{Marie and Pierre Curie with their daughter Ir\`ene. Marie Curie was awarded with Pierre and Becquerel the Nobel Prize for Physics in 1903, and then in 1911 the Nobel Prize for Chemistry (she is the only scientist awarded the Nobel prize for two different scientific disciplines). Ir\`ene will be awarded the  Nobel Prize for Chemistry in 1935 for her discovery of artificial radioactivity.}
\label{fig:curies}       
\end{figure}

In 1896 Becquerel \cite{Beq1896} discovered  spontaneous radioactivity. A few years later, Marie and Pierre Curie (Figure \ref{fig:curies}) discovered
\cite{Cur1898} that the elements Polonium and Radium suffered transmutations generating radioactivity: such transmutation processes were then called ``radioactive decays''. In the presence of a radioactive material, a charged electroscope prom\-ptly discharges. It was concluded that some elements were able to emit charged particles, that in turn caused the discharge of electroscopes.  The discharge rate of electroscopes was then used to gauge the level of radioactivity.

This observation opened in Europe and the New World (United States
and Canada in particular) a new era in research related to studies on natural radioactivity, and somehow unified, thanks to  the common experimental technique, studies of ionization in the context of meteorology and research related to natural radioactivity.



\begin{figure}
\begin{center}
\resizebox{0.94\columnwidth}{!}{\includegraphics{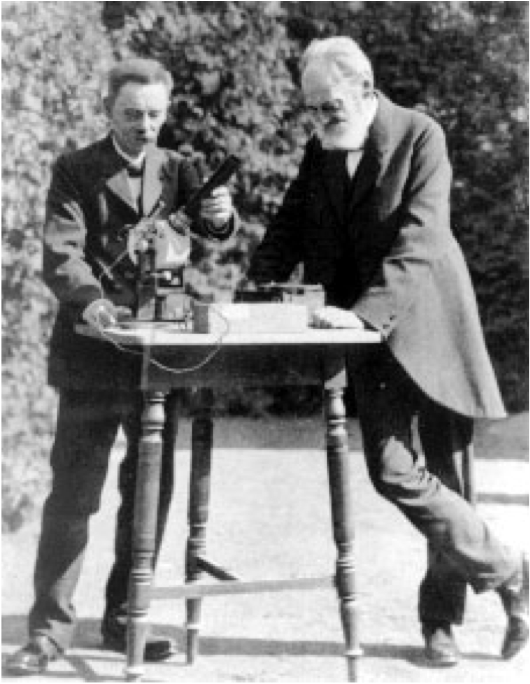} }
\end{center}
\caption{The two friends Julius Elster and Hans Geiter around 1900.}
\label{fig:elstergeitela}       
\end{figure}

\begin{figure}
\begin{center}
\resizebox{\columnwidth}{!}{\includegraphics{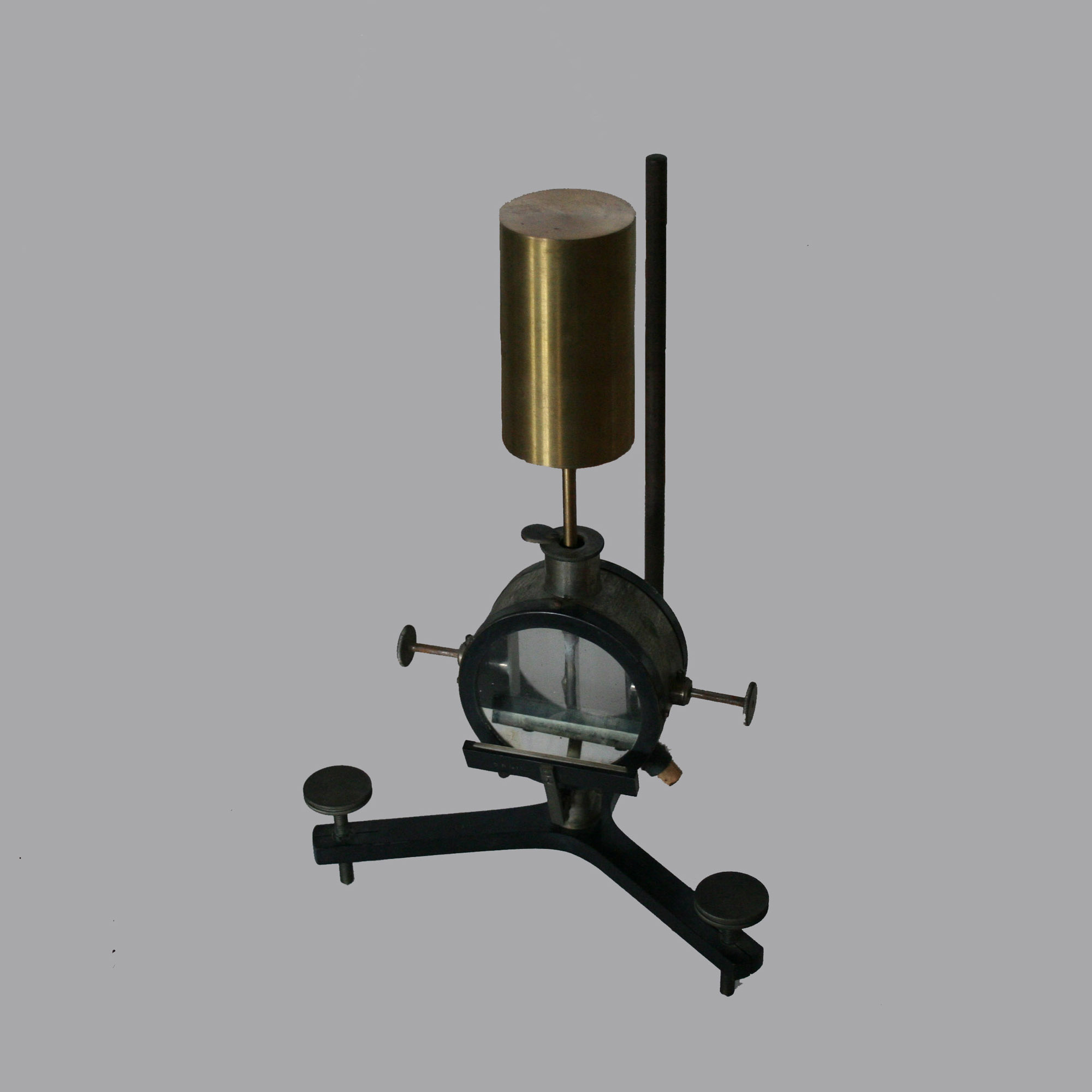} }
\end{center}
\caption{An electroscope developed by Elster and Geitel in the same period (private collection R. Fricke).}
\label{fig:elstergeitelb}       
\end{figure}

Around 1900,  Elster and Geitel (Figure \ref{fig:elstergeitela}) in Germany, and Wilson in England,  improved the technique for a
careful insulation of electroscopes in a closed vessel (Figure \ref{fig:elstergeitelb}), thus improving the sensitivity of
the electroscope itself. As a result, they could make quantitative measurements of the
rate of spontaneous discharge.

Julius Elster (1854-1920) and Hans Geitel (1855-1923) were two high school teachers in  Wolfenb\"uttel, a small town in Lower Saxony; friends since their time in school, they shared the same house with the family of Elster and worked maniacally to study the properties of electricity in the air. In a key experiment in 1899, they isolated the electroscope by putting it in a thick metal box. 
Also in these conditions they found a decrease in radioactivity, thus concluding \cite{ElG1900}
that the discharge was largely due to ionizing agents from outside the container. They also found that such ionizing agents were highly penetrating. The obvious question was if the radiation measured was coming from the ground, from the atmosphere, or if it was  extra-terrestrial. The simplest hypothesis was that its origin was linked to radioactive materials, and thus the terrestrial origin was the common assumption; however a demonstration seemed difficult to achieve.

\begin{figure}
\begin{center}
\resizebox{0.8\columnwidth}{!}{ \includegraphics{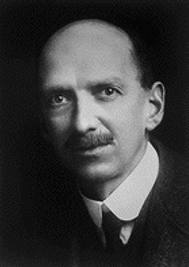} }
\end{center}
\caption{C.T.R. Wilson.}
\label{fig:wilsona}       
\end{figure}

\begin{figure}
\begin{center}
\resizebox{\columnwidth}{!}{ \includegraphics{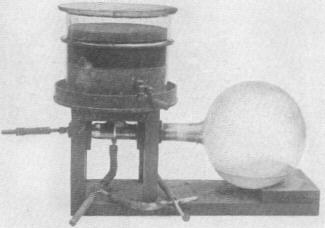} }
\end{center}
\caption{A cloud chamber built by Wilson.}
\label{fig:wilsonb}       
\end{figure}

Charles Thomson Rees Wilson (1869 -- 1959; Figure \ref{fig:wilsona}) was a Scottish physicist; in 1911 he will invent the detector called ``cloud chamber'' or ``Wilson chamber'' (Figure \ref{fig:wilsonb}),  which has been fundamental for the history of physics and of cosmic rays in particular -- in 1927 he will be awarded the Nobel prize for physics. Wilson was
interested to the problem of the origin of penetrating radiation, and  immediately in 1901 confirmed the result by Elster and Geitel,
and suggested the possibility that the source of the
ionization could be extraterrestrial \cite{Wil1901}.
He wrote: ``we must conduct experiments to determine if the production of
ions in the air free of impurities can be explained as arising from external sources, probably R\"ontgen radiation rays or cathode rays, but
largely more penetrating''. Wilson carried his electroscope in a tunnel in Scotland, screened
by the surrounding rock, but could not measure, because of experimental uncertainties, a reduction of radioactivity with respect to the open air as  he expected to find if the extraterrestrial hypothesis had been true. The
theory of an extraterrestrial origin of the radiation, although occasionally discussed,
was abandoned for the next ten years. However, the famous Serbian-Croatian-American engineer and inventor Nikola Tesla (1856 - 1943) patented in 1901 in the US a power generator based on the fact that ``the Sun, as well as other sources of radiant energy, throws off minute particles of matter [Éwhich] communicate an electrical charge". 

The results of Elster and Geitel, and Wilson, motivated in Germany and in England, in particular at the Cavendish Laboratory in Cambridge, a great interest in the question of the origin of natural radiation.
 Vienna was
also one of the focal points of this kind of research in Europe. Across the ocean, in Canada, during 1903
Rutherford  and Cooke and McLennan and Burton made several experiments by changing the conditions of
insulation of electroscopes, in particular by placing insulating walls in different directions. They concluded
(not without courage, given the experimental uncertainties) that the radiation seemed to
come from all directions with the same intensity.

In the period from  1906 to 1908, extensive systematic research was carried out 
to characterize the source of radiation. Fluctuations related to position, time of the day, pressure and temperature were larger than the precision of the instruments, and it seemed impossible to get a clear picture.
Strong in 1907 measured the radioactivity in many different places including his laboratory, the center of a tank filled with rainwater, and open air; the results
were dominated by experimental errors, and the fluctuations could not allow  drawing
conclusions.
Between 1907 and 1908, Eve made measurements over the Atlantic Ocean, which showed
within the errors comparable levels of radioactivity above  the sea, in England and in Montreal.

The result of the many experiments performed up to 1909 was that the
spontaneous
discharge was consistent with the hypothesis that even in insulated environments
a background radiation did exist. Calculations were made of how this radiation, thought to be gamma radiation due to its 
penetrating power, should decrease with
height (in particular by the Eve), and attenuation measurements were performed.

\section{Father Wulf}

\begin{figure}
\begin{center}
\resizebox{\columnwidth}{!}{\includegraphics{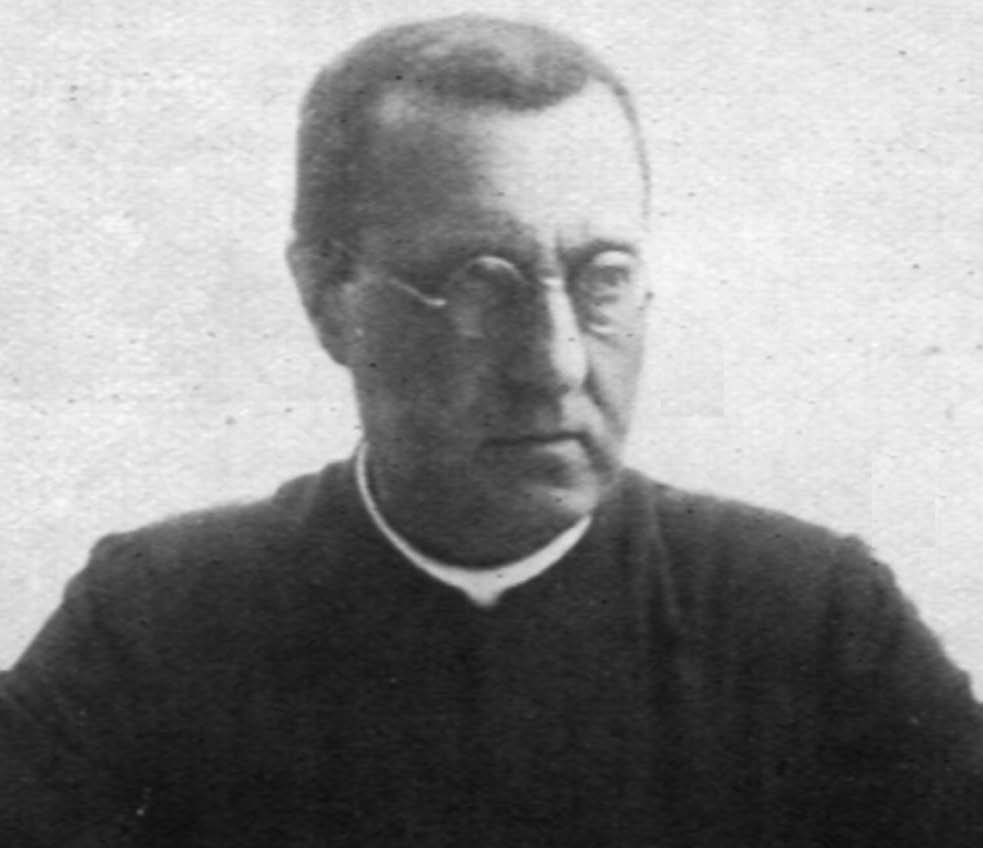} }
\end{center}
\caption{Father Theodor Wulf around 1910 (archive of the S.J., Munich).}
\label{fig:wulf}       
\end{figure}

\begin{figure}
\begin{center}
\resizebox{0.8\columnwidth}{!}{\includegraphics{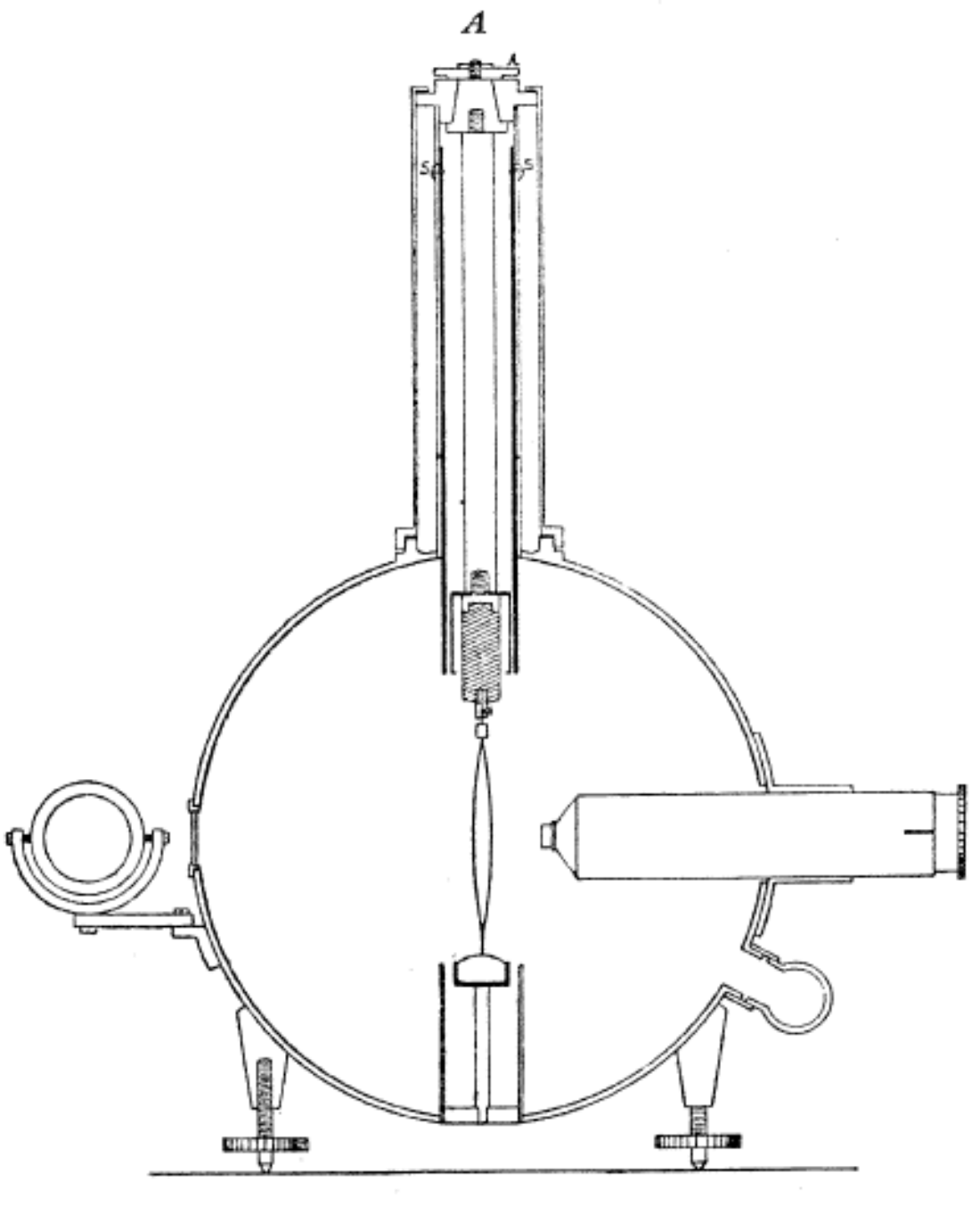} }
\end{center}
\caption{Scheme of the Wulf electroscope (drawn by Wulf himself). The 17 cm diameter cylinder with depth 13 cm was made
of Zinc. To the right is the microscope that measures the distance between the two silicon
glass wires illuminated using the mirror to the left. According to  Wulf, the sensitivity of the instrument,
as measured by the decrease of the inter-wire distance, was 1 volt.}
\label{fig:1aa}       
\end{figure}

\begin{figure}
\begin{center}
\resizebox{0.9\columnwidth}{!}{\includegraphics{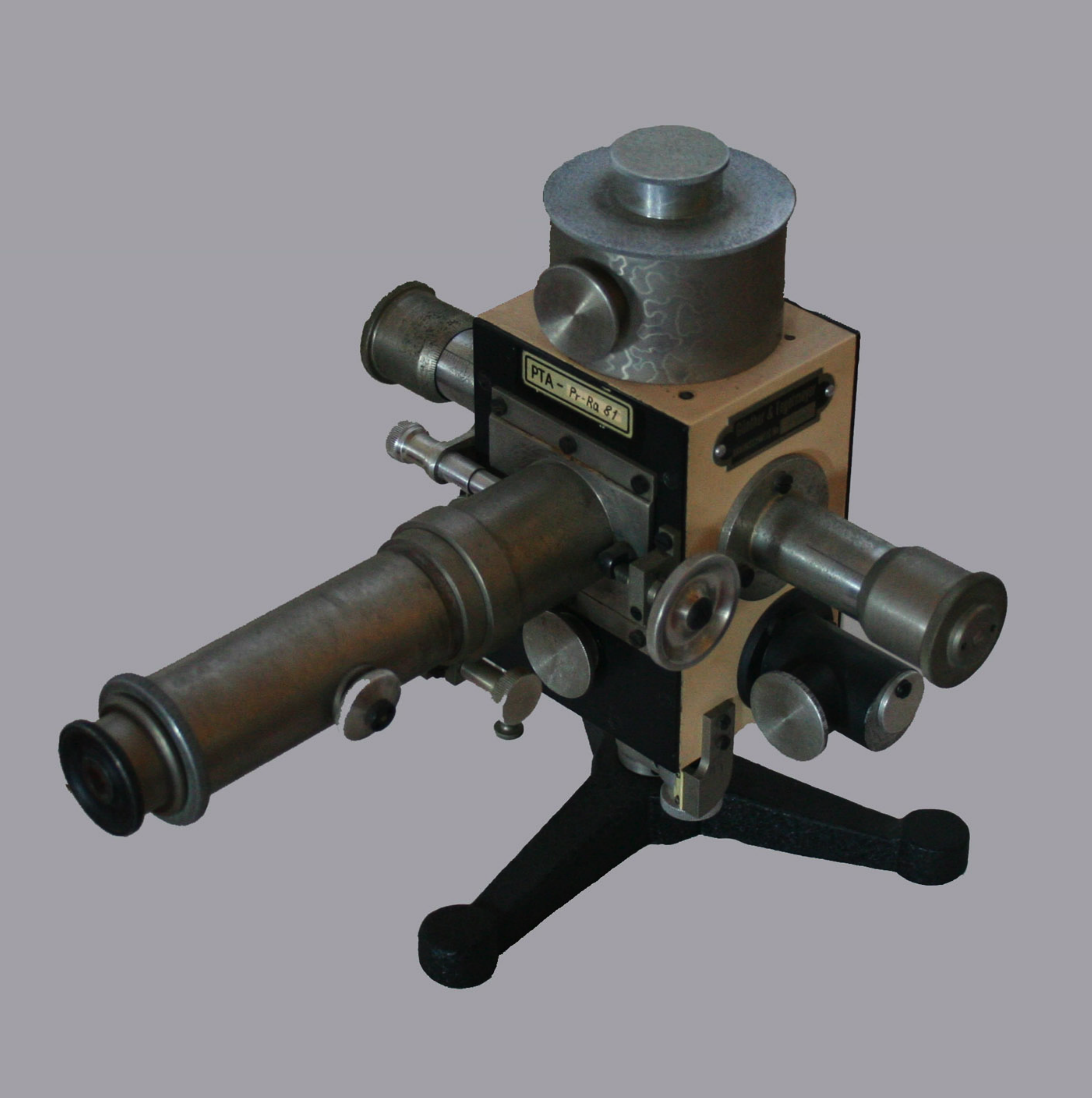} }
\end{center}
\caption{An electroscope user by Wulf (private collection R. Fricke).}
\label{fig:1ab}       
\end{figure}

The electroscope was a delicate instrument, difficult to transport: a technical improvement was needed
to make measurements easier. In addition, new ideas on what to look for in
measurements were possibly needed. Although the leading institute for the research on atmospheric ionization was Vienna,
these questions were answered by  the fundamental work of an independent researcher, father Wulf (Figure \ref{fig:wulf}).

Father Theodor Wulf (1868 - 1946) was a
German scientist; he became a Jesuit priest at the age of 20, before studying physics
under the direction of Walther Nernst at the University of G\"ottingen. He taught physics at the
Jesuit University of Valkenburg, in the Netherlands, from 1904 to 1914 and from 1918 to 1935, and worked at 
the Collegio Romano in Rome.

Wulf designed and built a more sensitive, and above all, more transportable, electrometer
than normal gold-leaf electroscopes: in Wulf's electroscope the two leaves
were replaced by two strips of metalised glass in tension.

%
%

In 1909, Wulf tested his electroscope by measuring the ionization in various locations in Germany, Holland and Belgium. He concluded that the results of his experiments confirmed the validity the instrument he had developed, and that everything was consistent with the hypothesis that the penetrating radiation was caused by radioactive substances present in upper layers of the Earth's crust. Temporal variations were  interpreted as caused by fluctuations in air pressure or air flow. He wrote then that if there was an additional component, it was too small to be measured with the available instrumentation.

Once commissioned his instrument and verified the validity of the measurements, Wulf had the idea to measure the variation of radioactivity with height in order to understand its origin. The idea was simple: if the radioactivity  was coming from Earth, it would decrease with height.

  In 1909 and 1910 \cite{Wul1910} he traveled to Paris bringing his electroscope (Figures \ref{fig:1aa},\ref{fig:1ab}) with him, and measured the ionization rate on the top of the Eiffel Tower (about 300 meters high). Supporting the hypothesis of the terrestrial origin of most of the radiation, he expected to find at the top less ionisation than on the ground. The rate of ionisation showed, however, too small a decrease to confirm the hypothesis. He concluded that, in comparison with the values on the ground, the intensity of radiation ``decreases at nearly 300 m [altitude] not even to half of its ground value''; while with the assumption that radiation emerges from the ground there would remain at the top of the tower ``just a few percent of the ground radiation'' \cite{Wul1910}. 
 
Wulf's observations were of great value, because he could take data at different hours of the day and for many days at the same place. For a long time, Wulf's data were considered as the most reliable source of information on the altitude effect in the penetrating radiation. However Wulf concluded that the most likely explanation of his puzzling result was still emission from the Earth's crust. Wulf's experiment was impressive due to its simplicity. As Einstein used to say, nature cannot be oversimplified: the failure of Wulf's experiment was due to the fact that the metal of which the Eiffel tower is made is radioactive, and to a subtle combination of decrease of the contribution by ground radioactivity and increase of the contribution of cosmic rays with height.  Wulf's technique paved the way for the final discovery, as we shall see.

%


The prevailing interpretation of all results in the beginning of the XX century was that penetrating radioactivity came mainly or completely from radioactive materials in the Earth's crust. In the 1909 review by Karl Kurz \cite{kurz} three possible sources for
the penetrating radiation are discussed: an extra-terrestrial radiation possibly from the
Sun; radioactivity from the crust of the Earth; and radioactivity in the atmosphere.
Kurz concludes, however, that the possibility of an extraterrestrial radiation seems unlikely. But a handful of independent researchers across Europe did not agree, and they were designing new experiments.

\section{Pacini and the measurements underwater}

The conclusion that radioactivity was mostly coming from the Earth's crust was questioned by the Italian physicist Domenico Pacini (1878 - 1934). Pacini graduated in Rome supervised by professor Pietro Blaserna, an Austrian-born physicist who had studied and worked in Vienna; he did most of his research work as an assistant meteorologist in Rome, and became later a professor of experimental physics in the University of Bari.

\begin{figure}
\begin{center}
\resizebox{\columnwidth}{!}{\includegraphics{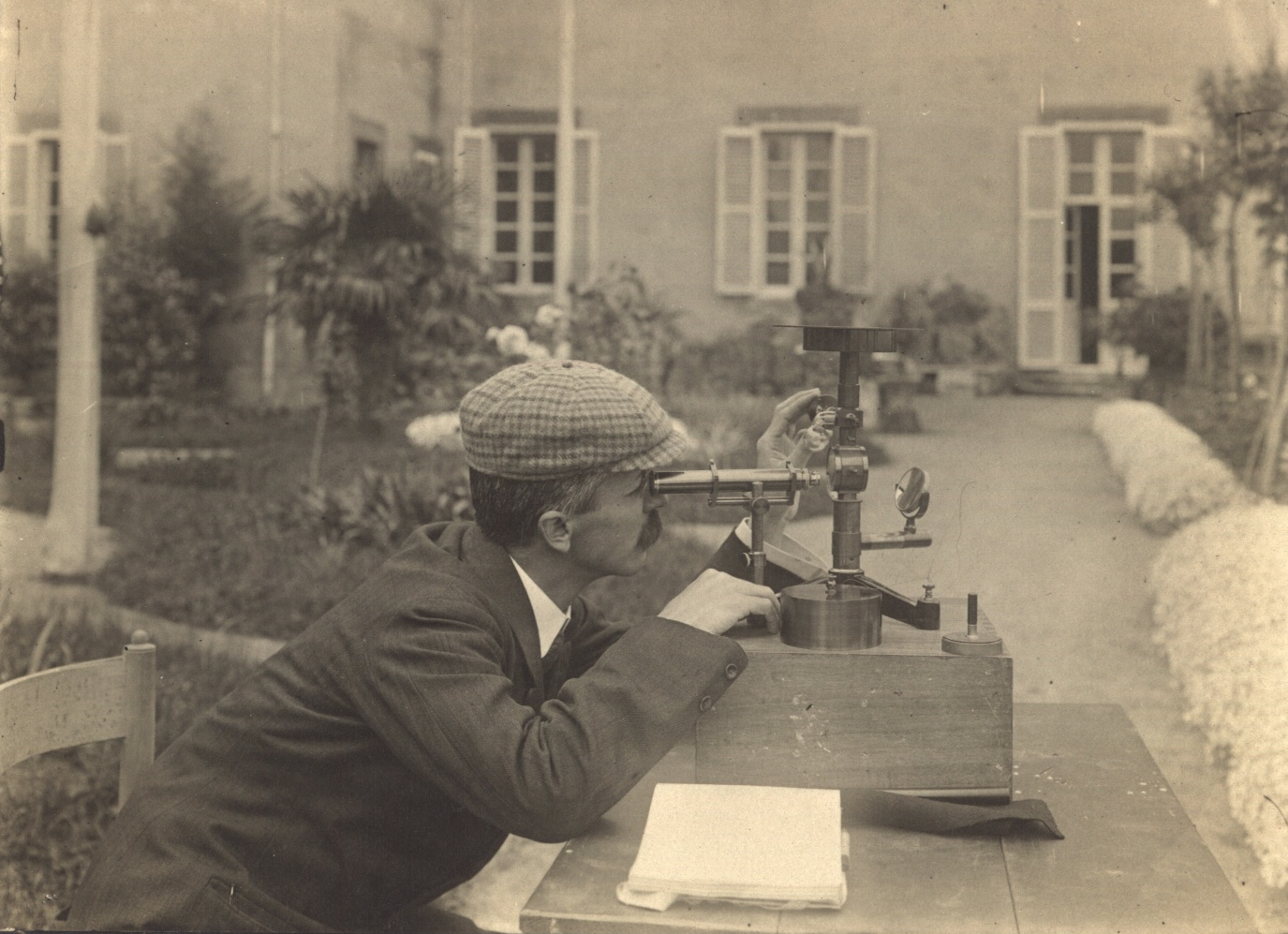} }
\end{center}
\caption{Pacini making a measurement in 1910 (courtesy of the Pacini family).}
\label{fig:pacini}       
\end{figure}

\begin{figure}
\begin{center}
\resizebox{\columnwidth}{!}{\includegraphics{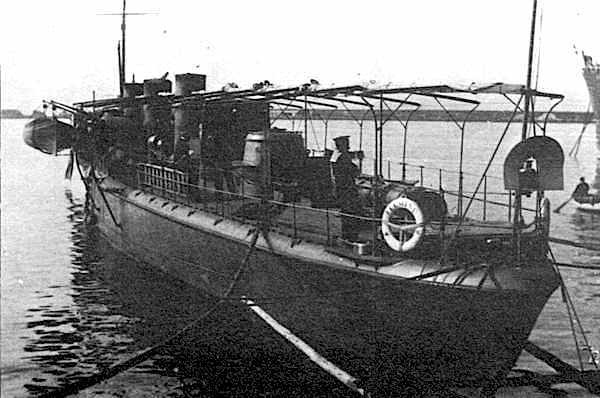} }
\end{center}
\caption{The cacciatorpediniere ``Fulmine'', used by Pacini for his measurements on the sea (courtesy of the Marina Militare Italiana).}
\label{fig:fulmine}       
\end{figure}

Pacini (Figure \ref{fig:pacini}) started then an experimental program of systematic measurements of radiation 
on the ground 
(at different elevations, including at sea level, and in different places to study local effects) and on the sea~\cite{Pac1909,Pac1910}. 
Those measurements  were aimed at checking 
whether the radioactivity within the Earth's crust was sufficient to explain the 
ionization effects (about 13 ions per second per cubic centimeter of air) that had been 
measured on the Earth's surface. 
In the paper \cite{Pac1910} Pacini measured  onboard the cacciatorpediniere (destroyer) ``Fulmine'' from the Italian Navy (Figure \ref{fig:fulmine}),  that the ionization on the sea surface, 300 m from the beach of Livorno (in front of the Naval Academy), was about two thirds of the ionization on the ground, 
thus supporting
 the idea that a non negligible part of the penetrating radiation  is independent of the emission from the Earth's crust.

The definitive experiment is, however, the one performed in June 1911 \cite{Pac1912}, during 7 days of measurements in the deep sea in the Genova gulf, in front of the Naval Academy of Livorno. This measurement has an important place in the history of physics, since it pioneers the technique of underwater measurement of radiation.

Pacini  developed  an experimental technique for underwater measurements. He found a significant decrease in the discharge rate when the electroscope was placed three meters underwater, first in the sea of the gulf of Genova, and then in the Lake of Bracciano. 
He wrote: ``Observations carried out on the sea during the year 1910  led me to conclude that a significant proportion of the pervasive radiation that is found in air had an origin that was independent of direct action of the active substances in the upper layers of the Earth's surface. ... [To prove this conclusion] the apparatus ... was enclosed in a copper box so that it could immerse in depth. [...] Observations were performed with the instrument at the surface, and with the instrument immersed in water, at a depth of 3 metres.''  
Pacini measured seven times during three hours
the discharge of the electroscope, measuring a ionization of
11.0 ions per cubic centimeter per second on surface; with the apparatus at a 3 m depth in the 7 m deep sea, he measured  8.9 ions per cubic centimeter  per second. The difference of 2.1 ions per cubic centimeter  per second (about 20\% of the total radiation) should be, in his view, attributed to an extraterrestrial radiation. The statistical significance of the difference was of 4.3 standard deviations.

The conclusion of his article, published in ``Nuovo Cimento'' in February 1912, were (the italics come from the original article) that  {\em``[it] appears from the results of the work described in this Note that a sizable cause of ionisation exists in the atmosphere, originating from penetrating radiation, independent of the direct action of radioactive substances in the soil." }

As a curiosity, in 1910 Pacini had looked for a possible increase in radioactivity during a passage of the Halley's comet \cite{PaciniHalley}, and he found no 
evidence of a measurable effect from the comet itself.

\section{The first balloon flyers}

The need for balloon experiments 
became evident to clarify Wulf's observations on the effect of altitude.
It must be said that  balloon experiments were anyway
widely used for studying atmospheric electricity, in particular in Vienna. 

\begin{figure}
\begin{center}
\resizebox{0.7\columnwidth}{!}{\includegraphics{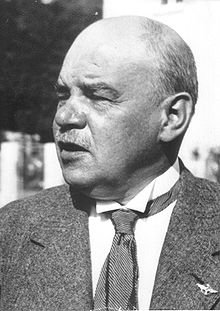} }
\end{center}
\caption{Franz Linke (1878 - 1944)-- source: wikimedia commons.}
\label{fig:linke}       
\end{figure}
\begin{figure}
\begin{center}
\resizebox{0.8\columnwidth}{!}{ \includegraphics{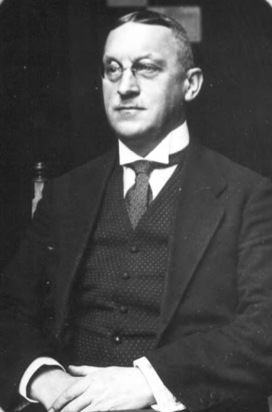} }
\end{center}
\caption{Karl Bergwitz.}
\label{fig:berg}       
\end{figure}

The fist balloon 
flight with the purpose of studying the properties
of penetrating radiation\footnote{The meteorologist Franz Linke (Figure \ref{fig:linke}) had, in fact, made 12 balloon flights between  1900 and 1903 during his PhD studies at Berlin University, carrying an electroscope built by Elster and Geitel to heights up to 5500 m. His measurements were not motivated by the study of radioactivity; however, his thesis \cite{linkethesis} concludes: ``Were one to compare the presented values with those on ground, one must say that at 1000 m altitude [...] the ionization is smaller than on the ground, between 1 and 3 km the same amount, and above it is larger, with values increasing up to a factor of 4 (at 5500 m). [...] The uncertainties in the observations [...] only allow the conclusion that the reason for the ionization has to be found first in the Earth.'' After his thesis Linke went to Asia; nobody later quoted him, but although he had made the right measurement, he had reached the wrong conclusions.} was arranged in 1909 by Karl Bergwitz (Figure \ref{fig:berg}), a former gymnasium pupil of Elster and Geitel. Bergwitz found that the ionization at 1300 m altitude had decreased to about 24\% of the value on the ground. However, Bergwitz's results were questioned because his electrometer was damaged during the flight. He later investigated his electrometers on the ground, and finally reported \cite{bergwitz}  that he had observed no significant decrease of the ionization. 

\begin{figure}
\begin{center}
\resizebox{0.8\columnwidth}{!}{ \includegraphics{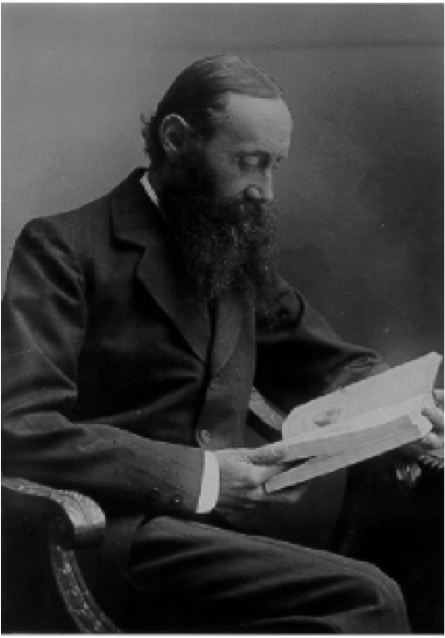} }
\end{center}
\caption{Albert Gockel.}
\label{fig:gockela}       
\end{figure}

\begin{figure}
\begin{center}
\resizebox{\columnwidth}{!}{ \includegraphics{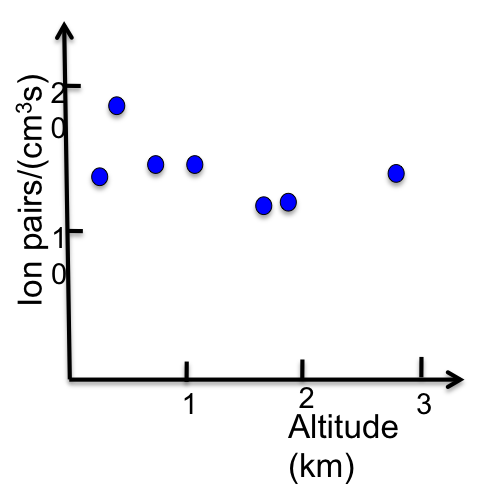} }
\end{center}
\caption{Ionization measured by Gockel at various altitudes.}
\label{fig:gockelb}       
\end{figure}

A few months later,  Albert Gockel (1860 - 1927, Figure \ref{fig:gockela}), professor at the University of
Fribourg, ascending up to 4500 m above sea level  and measuring up to 3000 m during three successive 
flights with a balloon
from the Swiss aeroclub,
found  \cite{gockel1910} that the ionization did not decrease with height (Figure \ref{fig:gockelb}) as expected in the hypothesis
of a terrestrial origin. Gockel confirmed the conclusions of Pacini, quoting
correctly his 1910 article, and concluded that ``a non-negligible part of the penetrating radiation is
independent of the direct action of the radioactive substances in the uppermost layers of
the Earth''. Gockel first introduced the term ``kosmische Strahlung''.
 
Indeed Gockel had been particularly unlucky. Subsequent calculations, carried out by
Schr\"odinger during her thesis work, showed that if the radioactivity comes in part from Earth and in part from above (as is the case), up to three thousand meters the
decrease of radioactivity from the Earth's crust can be offset by
growth of the radioactivity from extraterrestrial sources. Such a conclusion was reported
later by Hess -- who perhaps knew the Schr\"odinger calculations since both were in Vienna in the same group -- without quoting his younger colleague. If Gockel had insisted flying higher he could have obtained
a significant result. 

Despite the findings of Pacini, and the results (not conclusive) of Wulf and Gockel on the
dependence  of the radioactivity on  altitude, physicists were still reluctant to abandon the hypothesis of terrestrial origin.

The situation will be clarified through the famous long series of balloon flights by
Victor Hess, a masterpiece in the history of physics due to the careful preparation and the scrupulous execution of the experiment.

\section{The recognition of the early works by the scientific community}

Hess is today remembered as the discoverer of cosmic rays for which he was awarded the 1936 Nobel Prize in physics, nominated by Compton (Pacini died two years before, in 1934, and was thus not eligible). In his nomination Compton had written: ``The time has now arrived, it seems to me, when we can say that the so-called cosmic rays definitely have their origin at such remote distances from the Earth that they may properly be called cosmic, and that the use of the rays has by now led to results of such importance that they may be considered a discovery of the first magnitude. [...] It is, I believe, correct to say that Hess was the first to establish the increase of the ionization observed in electroscopes with increasing altitude; and he was certainly the first to ascribe with confidence this increased ionization to radiation coming from outside the Earth''. Why so late a recognition? Compton writes: ``Before it was appropriate to award the Nobel Prize for the discovery of these rays, it was necessary to await more positive evidence regarding their unique characteristics and importance in various fields of physics" \cite{noi2}.   The Nobel prize to Hess was shared with
C.D. Anderson for the discovery of the positron.

Hess' discovery was based on contributions of many other scientists. Elster and Geitel, and then Wulf, developed the instrument used for the measurement. Wilson first fromulated the hypothesis of a cosmic origin of radiation. Pacini concluded that part of the  ``penetrating radiation'' was extraterrestrial one year before Hess (the technique used by Pacini, anyway, could not firmly disprove a possible atmospheric origin of
the background radiation). Gockel and Linke flew at high altitude, and found results incompatible with a terrestrial origin of radiation.  The final report by the Nobel prize Committee to the Royal Academy of Sweden quotes Gockel's conclusion that the results of his balloon measurements, in agreement with measurements by Pacini, show that a not insignificant part of the radiation is independent of direct action of substances in the crust of the Earth;  it notes however that Hess' careful work includes an accurate measurement of the absorption of gamma rays, and several balloon ascents in 1911 and 1912, and that it shows for the first time that a very penetrating radiation is incident on the atmosphere from the outside \cite{noi2}.
%

In any case, a whole
community of researchers was involved in that field. Pacini certainly made a break-through by
introducing the technique of underwater measurement and by finding a significant decrease of the radiation with respect to the surface, which could exclude the Earth as the only source of radiation; 
unfortunately, he could not participate properly to the ongoing debate,
and he could not push his results with energy. Pacini's work
was carried out in difficult conditions because of lack of resources available
to him, because of lack of scientific freedom during the crucial years when he
was working at the Central Bureau of Meteorology and Geodynamics, and
because of the substantial indifference his work was met with by the
Italian academic world - that, {\em{per se,}} made Nobel Prize nominations difficult. 

Some excerpts from mail exchanges
 that occurred
between Hess and Pacini in 1920 \cite{Riz1934} are very illuminating. 

On March 6, 1920, Pacini wrote to Hess: ``...I had the opportunity to study some of your papers about electrical-atmospheric phenomena that you submitted to the Principal Director of
the Central Bureau of Meteorology and Geodynamics [in Rome]. I was already aware of some of these works from summaries that had been reported to me during the war. [But] the paper \cite{Hess1919} entitled  `The problem of penetrating radiation of extraterrestrial origin'\footnote{`Die Frage der durchdringenden Strahlung ausserterrestrischen Ursprunges'} was unknown to
me. While I have to congratulate you for the clarity in which this important matter is
explained, I have to remark, unfortunately, that the Italian measurements and observations,
which take priority as far as the conclusions that you, Gockel and Kolh\"orster
draw, are missing; and I am so sorry about this, because in my own publications I
never forgot to mention and cite anyone...''.

The answer by Hess, dated March 17, 1920, was: ``Dear Mr. Professor, your very
valuable letter dated March 6 was to me particularly precious because it gave me the
opportunity to re-establish our links that unfortunately were severed during the war.
I could have contacted you before, but unfortunately I did not know your address. My
short paper `The problem of penetrating radiation of extraterrestrial origin'
is a report of a public conference, and therefore has no claim of completeness. Since
it reported the first balloon measurements, I did not provide an in-depth explanation
of your sea measurements, which are well known to me. Therefore please excuse me
for my unkind omission, that was truly far from my aim ...''. 
On April 12, 1920,
Pacini in turn replied to Hess: ``... [W]hat you say about the measurements on the
penetrating radiation performed on balloon is correct; however the paper `The problem of penetrating radiation of extraterrestrial origin'  lingers quite a bit on
measurements of the attenuation of this radiation made before your balloon flights,
and several authors are cited whereas I do not see any reference to my relevant
measurements (on the same matter) performed underwater in the sea and in the
Bracciano Lake, that led me to the same conclusions that the balloon flights have
later confirmed.''

Finally, on May 20, 1920, Hess replied to Pacini: ``...Coming back to your publication in `Nuovo Cimento', (6) 3 Vol. 93, February 1912, I am ready to acknowledge that certainly you had the priority in expressing the statement, that a non terrestrial radiation of 2 ions/cm$^3$ per second at sea level is present.  However, the demonstration of
the existence of a new
source of penetrating radiation
from above
came from my balloon ascent to a height of 5000 meters on August 7 1912, in which I have discovered a huge increase in radiation above 3000 meters.''

The Hess-Pacini correspondence, nine years after Pacini's work and eight years after Hess' 1912 balloon flight, shows how difficult communication was at the time. Also language difficulties may have contributed: Pacini publishing mostly in Italian and Hess in German.

\section{Conclusion}

The work behind the discovery of cosmic rays, a milestone in science, comprised scientists in Europe, Canada, 
and the US, and took place during a period characterized by lack of communication and by nationalism caused primarily by the World War I.
In the work that culminated with high altitude balloon flights by Hess,  important contributions have been forgotten and in particular those of Elster and Geitel, Wilson, Wulf, Linke, Bergwitz, Gockel, Pacini. Besides the personal stories related to the accidents of the lives of individuals, several historical and political facts contributed to the lack of references to some important contributions to the discovery of cosmic rays, and should serve as a lesson for the
scientific politics of the future. 

\paragraph{\bf{Acknowledgements}} I thank Per Carlson. Most reflections on the role of nationalism in science come from discussion with him -- and discussions with him are always a great pleasure and a source of enrichment.

\end{document}